\documentclass[letterpaper, twocolumn, superscriptaddress, showpacs, preprintnumbers, amsmath, amssymb]{revtex4}
\usepackage{graphicx}
\usepackage{soul}
\usepackage{dcolumn}
\usepackage{bm}
\usepackage{color}
\usepackage{hyperref}
\usepackage{caption}
\usepackage[lofdepth,lotdepth]{subfig}

\def\be{\begin{equation}}
\def\ee{\end{equation}}
\def\bea{\begin{eqnarray}}
\def\eea{\end{eqnarray}}

\begin{document}

\title{Driven nonlinear dynamics of two coupled exchange-only qubits}
\author{Arijeet Pal, Emmanuel I. Rashba, and Bertrand I. Halperin}
\affiliation{Lyman Laboratory of Physics, Harvard University, Cambridge, MA 02138, USA}

\begin{abstract}
Inspired by creation of a fast exchange-only qubit (Medford et al., Phys. Rev. Lett., {\bf 111}, 050501 (2013)), we develop a theory describing 
the nonlinear dynamics of two such qubits that are capacitively coupled, when one of them is driven resonantly 
  at a frequency equal to its level splitting.  We include conditions of strong driving, where the Rabi frequency is a 
significant fraction of the level splitting, and we consider situations where the splitting for the second qubit may be 
the same or different than the first.  We demonstrate that coupling between qubits can be 
detected by reading the response of the second qubit, even when the coupling between them is only of about 1\% of their level splittings, and calculate entanglement between qubits. Patterns of nonlinear dynamics of coupled qubits and their entanglement are strongly dependent 
on the geometry of the system, and the specific mechanism of inter-qubit coupling deeply influences 
dynamics of both qubits. In particular, we describe the development of irregular dynamics in a two-qubit system, 
explore approaches for inhibiting it, and demonstrate existence of an optimal range of coupling strength 
maintaining stability during the operational time.
\end{abstract}

\maketitle

\section{Introduction}
\label{sec:intro}

Electrically gated quantum dots provide a promising platform for realizing qubits serving as building blocks of a quantum computer \cite{Loss,Levy,NC, Hanson, Kloeffel}. Two-level systems created from one or a few electrons in quantum dots have been experimentally isolated, initialized, and coherently manipulated. Such qubits were fabricated from single \cite{Koppens, Tokura, Nowack1, Brunner, Berg}, double \cite{Petta, Shulman, SNC, Higginb}, and triple quantum dot \cite{Laird,Sachrajda} structures. Reduction of decoherence is achieved by applying dynamical decoupling techniques \cite{Wang, Viola, Bluhm}. All techniques based on a single quantum dot or double quantum dots require, for performing two-axis rotations of the electron spin (or a pseudospin) on the Bloch sphere either high-frequency magnetic fields, or magnetic field gradients (from micromagnets or dynamical nuclear polarization), or spin-orbit coupling. Employing triple-dot qubits allows performing two-axes rotations by using the Heisenberg exchange only and completely by electrical means, which allows achieving fast performance by applying voltages to the gates.  

Such an approach based on coded qubits with the total electron spin $S=1/2$ and its projection $S_z=1/2$ was proposed by DiVincenzo et al. \cite{DiVincenzo} and realized experimentally by Laird et al. \cite{Laird}. More recently, two-axis rotations in such exchange-only qubits and their tomographic description were achieved \cite{Medford1,Aers_note}. A specific property of these qubits, requiring nonstandard techniques for their operation, stems from the fact that their natural rotation axes are not mutually perpendicular but intersect at an angle $2\pi/3$ (See Fig.~1(a)). This is also the case for three-electron double-dot qubits of Ref.~\cite{SNC}. For single-qubit operations this problem was resolved in Ref.~\cite{Medford2} by tuning the qubit into the resonant exchange regime, which is defined in Sec.~\ref{sec:exchange} below. Operations performed on such a resonant exchange (RX) qubit are very fast, at a scale of a few nanoseconds, with the Rabi nutation frequency comparable to the qubit level splitting. With the relaxation times $T_1\sim 40 \mu$s and $T_2\sim 1 \mu$s (even  without applying CPMG pulse sequences) \cite{Medford2} this qubit is a highly promising candidate for developing few-qubit systems. We take advantage of this separation of time scales between the operational and relaxation times and concentrate on the coupled dynamics of two qubits. For a reasonable magnitude of inter-qubit coupling, the duration of two-qubit operations is within the window of coherent operation. 
\begin{figure}[!hbtp]
\includegraphics[width=3.2in]{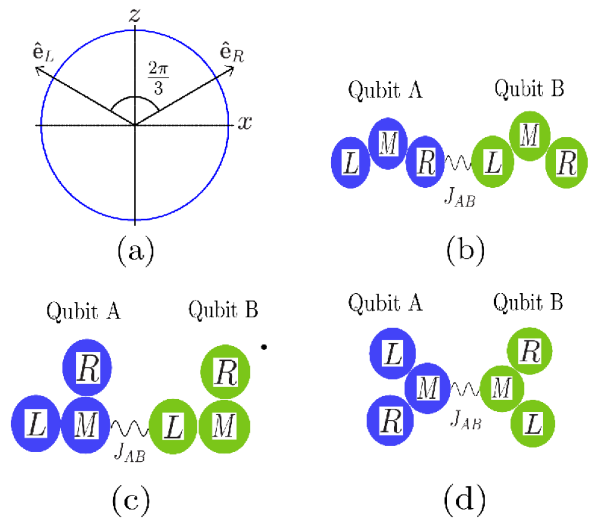}
\caption{(a) Bloch sphere of the exchange-only qubit. $\mathbf{\hat{e}}_L$ and $\mathbf{\hat{e}}_R$ are unit vectors for the left and right rotation axes. They are eigenvectors of the Hamiltonian of Eq.~(\ref{eq1}) with $t_L=0$ and $t_R=0$, respectively. (b) R-L geometry: nearest neighbor capacitive interaction between right dot of qubit A and left dot of qubit B (c) M-L geometry: coupling between middle dot of qubit A and left dot of qubit B. (d) M-M geometry: coupling between middle dots of qubits A and B. Blue and green dots represent qubits $A$ and $B$, respectively. $J_{AB}$ is the inter-qubit coupling constant.}
\label{fig:qubit_geometry}
\end{figure}

Similarly to single- and double-dot qubits \cite{Brunner, Nowack, Shulman}, the next step is achieving two-qubit entanglement, and first proposals for capacitively \cite{Taylor} and exchange \cite{Doherty} coupled double RX qubits have been already made.  Establishing entanglement between two qubits is a demanding task, and in this paper we propose and pursue in depth a protocol based on resonant driving of one of the qubits (qubit $A$) at its level splitting and measuring the response of the second qubit (qubit $B$) to the electrical signal produced by qubit $A$. Therefore, qubit $B$ serves as a detector of coupling between the two qubits. In the discussion below we examine several models of capacitive coupling between qubits. This coupling, in the absence of external mechanisms, keeps both qubits inside their logical spaces.

First, our analysis shows that not only the magnitude of the signal induced in qubit $B$, but also gross pattern of its dynamics, depend critically on the specific geometry of inter-qubit coupling and its strength. There is an optimal range of the coupling strength because at stronger coupling (albeit rather modest) dynamics of the two-qubit system might become irregular. Second, the back-action of qubit $B$ onto qubit $A$, an essential part of entanglement,  influences profoundly its own dynamics. Third, the crossed axes geometry of both qubits, while not critical for the single-qubit operation, affects significantly the coupling between qubits and two-qubit dynamics. In particular, specific choice of the mechanism  of inter-qubit coupling, Fig.~1(b,c,d), is critical for the spectrum of frequencies dominating the dynamics of coupled qubits and avoiding early switching to irregular regime. While our analysis was performed for capacitive coupling, it has implications also for exchange coupling between qubits.

In what follows, we apply our protocol to three different schemes of capacitive coupling between two exchange-only qubits and demonstrate spectacular differences in their dynamics.

\section{Single exchange-only qubit}
\label{sec:exchange}

An exchange-only qubit is a three-electron triple quantum dot with Zeeman split energy levels operated in the regime where total electronic spin $S=1/2$ and its projection onto the magnetic field $S_z=1/2$ \cite{DiVincenzo,Laird}. Its resonant exchange modification is operated in the parameter range where the two-dimensional qubit space is well separated from the other states of the system including all $S=3/2$ states and $S=1/2, S_z=-1/2$ states \cite{Medford2}.

In what follows, the three dots of each of the qubits are designated as $L$, $M$, and $R$, for the left, middle, and right dots. As outlined in more detail in Appendix A, basic properties of the qubit can be described in the framework of a Hubbard model whose basis includes two states, (201) and (102), with inhomogeneous charge distributions, and two ``neutral" (111) states; numbers show populations of $(L,M,R)$ dots. An electron pair in a doubly occupied dot is in a singlet state since triplet states are essentially higher in energy. The effective Schroedinger equation for this Hubbard model is for a four-spinor $(v_L,v_0,v_1,v_R)^T$ defined by Eq.~(\ref{eq:wavefn}) of Appendix~\ref{appendix:Hq_deriv}; here $T$ stands for transpose, .

Exchange-only qubits operate predominantly in the (111) region, with (201) and (102) serving mostly for initialization and projective measurements. However, admixture of these states is critical for operation of the qubit and the inter-qubit coupling. In the center of the (111) region the ground state of the qubit is $|0\rangle \approx \frac{1}{\sqrt{6}} \left( | \uparrow \uparrow \downarrow \rangle + |\downarrow \uparrow \uparrow \rangle - 2|\uparrow \downarrow \uparrow \rangle\right)$ and its excited state is $|1\rangle \approx \frac{1}{\sqrt{2}} \left( |\uparrow \uparrow \downarrow \rangle - |\downarrow \uparrow \uparrow \rangle \right)$ when written in terms of the electron spins on dots $L$, $M$, and $R$. Exact qubit states include small admixtures of (201) and (102). Resonant exchange qubits operate near the center of the $(111)$ region where the (201) and (102) components are small and can be projected out. Then the effective $2\times2$ qubit Hamiltonian acting in the $(|0\rangle,|1\rangle)$ space is
\be
{\hat H}_q=-\frac{J_z}{2} \sigma_z-\frac{J_x}{2} \sigma_x,
\label{eq1}
\ee  
where $(\sigma_z,\sigma_x)$ are Pauli matrices of the qubit pseudospin (for brevity, spin in what follows) acting in the qubit space. The exchange integrals are
\be
J_z=\frac{t_L^2}{U+V}+\frac{t_R^2}{U-V},\,\,J_x=\sqrt{3}\left(\frac{t_{R}^2}{U-V}-\frac{t_{L}^2}{U+V}\right),
\label{eq2}
\ee
where the tunneling matrix elements $t_L$ and $t_R$ connect the (111) sector to (201) and (102) states, respectively, $U>0$ is a single site repulsion energy, and $V$ and $-V$ are potentials applied to the left and right dot. The $V$ terms in denominators reflect the effect of $|201\rangle$ and $|102\rangle$ components of the spinor. Equations (\ref{eq1}) and (\ref{eq2}) are applicable when $t_L/U,t_R/U\ll1$; see Appendix A for details.

In our discussion we set $\hbar=1$, and we refer to $J_z$ as the qubit level splitting. The resonant exchange qubit operates in the regime $t_L=t_R$ near the point $V=0$ where $J_x=0$ and $J_z$ is stationary, $dJ_z/dV=0$. For simplicity, we disregard the dependence of $t_L$ and $t_R$ on $V$. As seen from Eq.~(\ref{eq1}), matrices $(\sigma_x,\sigma_z)$ perform single-qubit rotations about the axes orthogonal in the qubit space. 

\section{Inter-qubit coupling}
\label{sec:TQC}

Assuming the qubits are operated in a regime when tunneling between qubits $A$ and $B$ is negligible, the qubit-qubit interaction is capacitive through the electrostatic charges. The dots in the qubits $A$ and $B$ are arranged in such a way that exchange coupling inside each of the qubits exists only between adjacent dots, but we consider three geometries shown in Fig.~1(a,b,c). They differ in the capacitive coupling between qubits, and in all cases we consider only the capacitive coupling between two adjacent dots belonging to different qubits. This provides us with three models that show rather different two-qubit dynamics as shown in Sec.~\ref{sec:DE} below. In each of the geometries, we use symbols $L$, $M$, and $R$ to numerate dots inside qubits.

We begin with a linear array of Fig.~1(b) where the rightmost dot of $A$ is proximate to the leftmost dot of $B$. The excess electron densities $\delta n_\alpha$ at the edges of these qubits are equal to $|v^A_R|^2$ and $|v_L^B|^2$ and can be conveniently expressed in terms of the amplitudes ($v_0$,  $v_1$) of the corresponding qubits using first and fourth rows of Eq.~(\ref{eq:Hmatrix}). In terms of the unit vectors $\mathbf{\hat{e}}_L$ and $\mathbf{\hat{e}}_R$ of Fig.~1(a), defined as ${\hat{\bf e}}_L=(-\sqrt{3}/2,1/2)$ and ${\hat{\bf e}}_R=(\sqrt{3}/2,1/2)$ by their Cartesian coordinates in the $xz$-plane, the excess electron densities on the edge dots are 
\begin{equation}
\label{eq:nexc}
\delta n_\alpha= 2 \frac{t_{\alpha}^2}{U^2} |(\mathbf{\hat{e}_{\alpha}} \cdot {\bf{v}})|^2, \quad {\bf{v}} =  \begin{pmatrix} v_0 \\ v_1 \end{pmatrix},
\end{equation}
where $\alpha=\text{L, R}$. Due to neutrality, $\delta n_M=-(\delta n_L+\delta n_R)$. Hence, charges on the outermost dots are proprtional to squares of projections of spin wave functions
onto the tilted axes. Even though the single-qubit gate operations are immune to the non-orthogonality of $({\hat{\bf e}}_L,{\hat{\bf e}}_R)$ axes,
the electric charges, and therefore the interaction between qubits, is affected by it. 
The qubit-qubit interaction expressed in terms of the wavefunction amplitudes of qubits $A$ and $B$ when it is dominated by excess charges on right dot of $A$ and left dot of $B$, Fig.~1(b), is
\begin{align}
\label{eq:Hint}
H_{\text{int}} = \frac{e^2}{\kappa R} \delta n_R^A \delta n_L^B=4J_{AB}|(\mathbf{\hat{e}}_R \cdot {\bf{v}}^A)|^2 |( \mathbf{\hat{e}}_L \cdot {\bf{v}}^B)|^2,  
\end{align}
where $J_{AB}=(e^2/\kappa R)(t^A_Rt^B_L/U^2)^2$. Here $\kappa$ is the dielectric constant and $R$ is the distance between neighboring end dots of the qubits A and B. Having in mind a complicated geometry of the system and patterns of screening, $R$ can be also considered as an effective coupling constant that should be measured experimentally.

Equation~(\ref{eq:Hint}) includes both renormalization corrections to single-qubit Hamiltonians and inter-qubit correlations. Omitting the former contributions as described in Appendix~\ref{appendix:Hq_int}, we arrive at the interaction 
Hamiltonian in terms of the qubit spin vector-matrices $\mbox{\boldmath$\sigma$}^A=(\sigma_x^A, \sigma_y^A, \sigma_z^A)$ and similarly for $\mbox{\boldmath$\sigma$}^B$
\begin{align}
{\hat H}_{\rm int}=J_{AB}
({\bf{\hat e}}_R\cdot\mbox{\boldmath$\sigma$}^A)(\mathbf{\hat{e}}_L \cdot \mbox{\boldmath$\sigma$}^B).
\label{eqRL}
\end{align}
We note that ${\hat H}_{\rm int}$ includes $\sigma_x$ and $\sigma_z$ matrices of both qubits.

In the geometry of Fig.~1(c), $M$ dot of qubit $A$ is capacitively coupled to $L$ dot of qubit $B$. The density $\delta n^A_M$ contributes to the correlation Hamiltonian only a factor proportional to $\sigma_z^A$ because $\hat{\bf e}_L+\hat{\bf e}_R=\hat{\bf z}$, see Eqs.~(\ref{eq:QexcessL}) and~(\ref{eq:QexcessR}) of Appendix~\ref{appendix:Hq_int}. Hence,
\be
{\hat H}_{\rm int}=-J_{AB}\sigma_z^A  (\mathbf{\hat{e}}_L \cdot \mbox{\boldmath$\sigma$}^B). 
\label{eqML}
\ee

There is one more geometry, see Fig.~1(d), where two qubits are coupled through their $M$ dots. Its effective Hamiltonian is
\be
{\hat H}_{\rm int}=J_{AB}\sigma_z^A\sigma_z^B.
\label{eqMM}
\ee
The implementation of the CPhase gate using the Hamiltonian in Eq.~\ref{eqMM} can be achieved by following standard protocols \cite{Doherty}.

Three models of Eqs.~(\ref{eqRL})-(\ref{eqMM}) are based on the assumption of capacitive coupling only between the two closest dots of $A$ and $B$ qubits. The experimental fact that sensor dots respond predominantly to charges on the qubit dots closest to them can serve as a justification for using these models. From the theoretical perspective, this regularity can be ascribed to screening of the long range Coulomb interaction by metallic gates. We expect that the drastic difference in spin dynamics of these models unveiled in Sec.~\ref{sec:DE} below will serve to choose optimal geometries of two-qubit systems.

\section{Dynamics and Entanglement}
\label{sec:DE}

We consider a protocol when both qubits are tuned to $J_x^A=J_x^B=0$, with an additional oscillating term $J_x^A/2=\epsilon\cos(\omega t+\phi)$ produced by an ac voltage applied between the outermost dots of qubit $A$ beginning at time $t=0$. Qubit $B$ responds only to a capacitive signal produced by qubit $A$, and therefore dynamics of qubit $B$ is a direct indication of the coupling between $A$ and $B$. 

To get an outlook at the signal produced by $A$, we choose $\omega=J_z^A$. If the ac perturbation is small, we can solve the problem analytically in the rotating wave approximation (RWA) with $|\psi_A\rangle_{\text{rot}} = \exp \left(-i\sigma_z^A \omega t/2 \right) |\psi_A\rangle$ \cite{Vandersypen}. Subsequently, we transform back to the lab-frame and all the following equations are presented in this frame.

In the zeroth order in $J_{AB}$, the solution for qubit A starting at the North pole of the Bloch sphere at the instant $t=0$ is
\be
{\bf v}^A(t)=\left(\begin{array}{c}e^{i\omega t/2}\cos(\epsilon t/2)\\
ie^{-i\omega t/2}e^{-i\phi}\sin(\epsilon t/2)
\end{array}\right).
\label{RWS}
\ee
The components of the spin ${\bf S}^A$ of qubit $A$ on its Bloch sphere are equal to the mean values of the components of the $\mbox{\boldmath$\sigma$}^A$ matrix, ${\bf S}^A=\langle\mbox{\boldmath$\sigma$}^A\rangle$. Projections of the spin along the $({\hat{\bf e}}_L,{\hat{\bf e}}_R)$ axes give the electron densities entering the interaction Hamiltonian according to Eqs.~(\ref{eqRL}) - (\ref{eqMM})
\bea
S_x^A&=&v_0^Av_1^{A*}+v_0^{A*}v_1^A,\nonumber\\
S_y^A&=&i(v_0^Av_1^{A*}-v_0^{A*}v_1^A),\nonumber\\
S_z^A&=&v_0^Av_0^{A*}-v_1^Av_1^{A*}.
\label{eqS}
\eea
Calculation based on Eq.~(\ref{RWS}) results in
\bea
S_x^A(t)&=&\sin\epsilon t\sin{(\omega t+\phi)},\label{eqSx}\\
S_y^A(t)&=&\sin\epsilon t\cos{(\omega t+\phi)},\label{eqSy}\\
S_z^A(t)&=&\cos\epsilon t.
\label{eqSz}
\eea

Now, we evaluate the response of qubit B due to the coupling $J_{AB}$ at first order. Comparison of Eq.~(\ref{eqSz}) with Eqs.~(\ref{eqRL})-(\ref{eqMM}) demonstrates a critical effect of the inter-qubit coupling mechanism onto the electric signal driving the qubit $B$. In the geometry of Fig.~1(b), the spectrum of the signal includes three frequencies: two Raman frequencies $\omega\pm\epsilon$ coming from Eqs.~(\ref{eqSx})-(\ref{eqSy}) and the Rabi frequency $\epsilon$ coming from Eq.~(\ref{eqSz}). In the geometries of Fig.~1(c,d) only the Rabi frequency is present. The dynamics of qubit $A$ at frequencies that differ from the driving frequency $\omega$ stems from the fact that qubits are anharmonic systems and was already observed in double-dot qubits \cite{LairdEDSR,Rashbahalffreq}.

It is a special property of the geometry of Fig.~1(d) that qubits are coupled through their $M$ dots, and therefore the ac perturbation experienced by qubit $B$ is symmetric rather than antisymmetric in its $L$ and $R$ dots. When qubit $B$ is tuned to the point $J_x^B=0$, its effective Hamiltonian is ${\hat H}_q^B=(-J_z^B/2+\epsilon_B\cos \epsilon t)\sigma_z^B$, where the time dependent term is a perturbation exerted by qubit $A$ in the RWA (with $\epsilon_B\sim J_{AB}$), and for simplicity we have chosen the phase $\phi_B=0$.  Because ${\hat H}_q^B$ commutes with $\sigma_z^B$, the $z$-projection of the qubit spin ${\bf S}^B$ is conserved, $S^B_z=$const. Solving the Schroedinger equation with the initial condition $S_x^B(0)=1, S_y^B=0$, and applying Eq.~(\ref{eqS}), we arrive at
\bea
S_x^B(t)&=&\cos(J_z^Bt-2\frac{\epsilon_B}{\epsilon}\sin \epsilon t),\nonumber\\
S_y^B(t)&=&-\sin(J_z^Bt-2\frac{\epsilon_B}{\epsilon}\sin \epsilon t).
\label{eqSMM}
\eea 
Therefore, ${\bf S}^B$ precesses about the $z$-axis with a phase modulated by the inter-qubit coupling $J_{AB}$.

In the experiments of Ref.~[\onlinecite{Medford2}] the Rabi frequency $\epsilon$ was large and comparable to the qubit level splitting $J_z^A$. Under such conditions the accuracy of RWA is reduced, and Eqs.~(\ref{eqSx})-(\ref{eqSz}) provide only a qualitative outlook onto the dynamics of qubit $A$. Therefore, in our simulations presented below we solve for the dynamics of qubit $A$ exactly. These simulations are necessary also to account for the back-action of qubit $B$ onto qubit $A$ that provides the entanglement mechanism. 

The magnitude of $J_{AB}$ is highly sensitive to the ratio $t/U$; we assume $t_L=t_R=t$. If to estimate $t/U$ from $J_z^A\approx 2t^2/U$ with $t\approx 17 \mu$eV \cite{Medford2}, then $t/U\approx4\times10^{-2}$, with the qubit level splitting $\approx 1.4 \mu$eV. $J_{AB}$ estimated with $\kappa\approx10$ and $R\approx200$ nm is about $J_{AB}\approx10^{-3}J_z^A$. However, different estimates result in larger values of $t/U$ \cite{MedfordPr}, and our simulations were performed for $J_{AB}/J_z^A$ of about $10^{-2}$ which seems realistic. In practice, however, the coupling $J_{AB}$ should be obtained from experiment. An important purpose of the protocols proposed in the paper is to enable a measurement of this coupling.
\begin{figure*}[!hbtp]
\hspace{0.3in}
\subfloat[]{
\includegraphics[width=2.5in, height=2.1in]{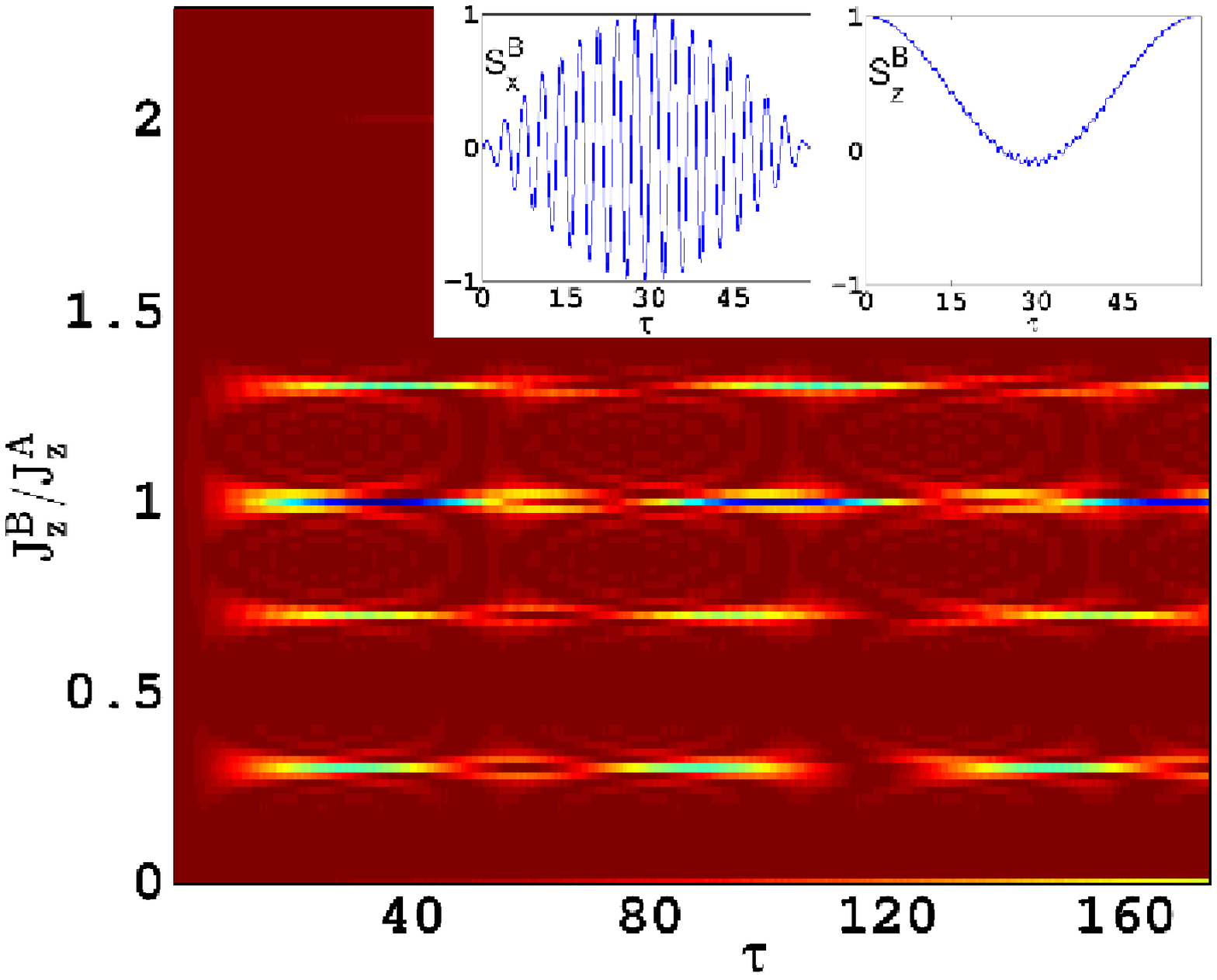}
\label{fig:RL_B0005}
}
\subfloat[]{
\includegraphics[width=2.75in, height=2.05in]{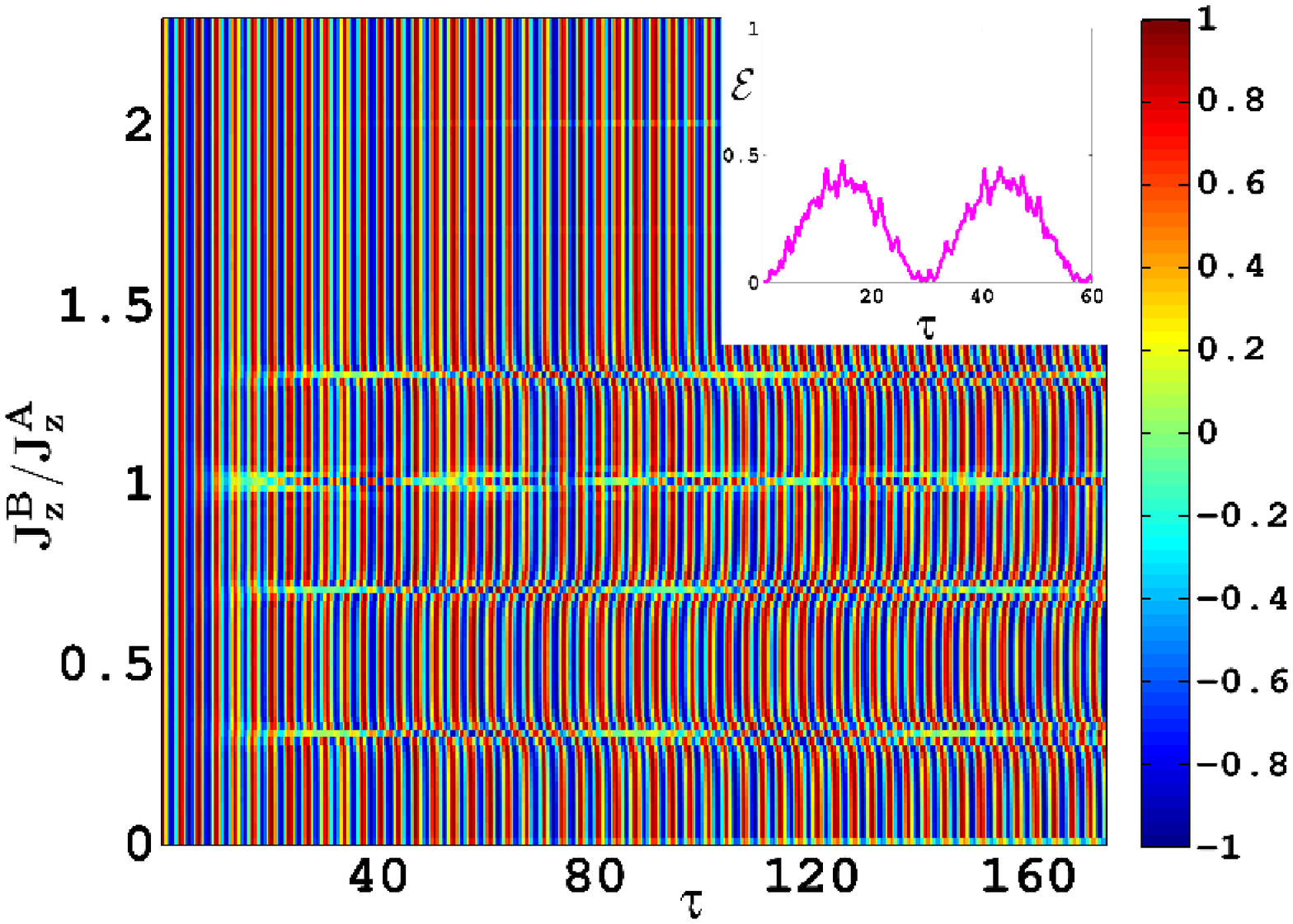}
\label{fig:RL_A0005}
}
\\
\centering
\subfloat[]{
\includegraphics[width=2.5in, height=2.1in]{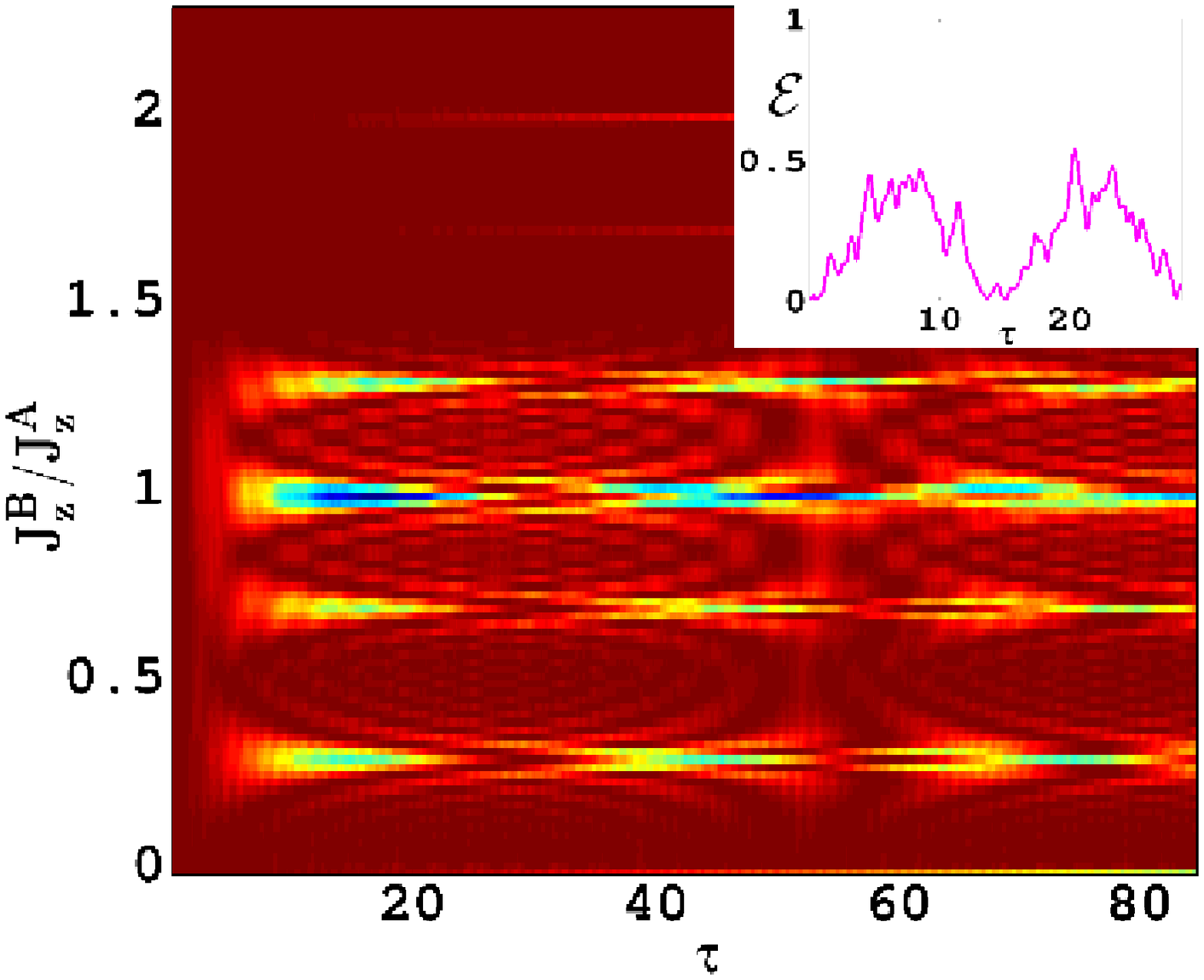}
\label{fig:RL_B001}
}
\subfloat[]{
\includegraphics[width=2.5in, height=2.1in]{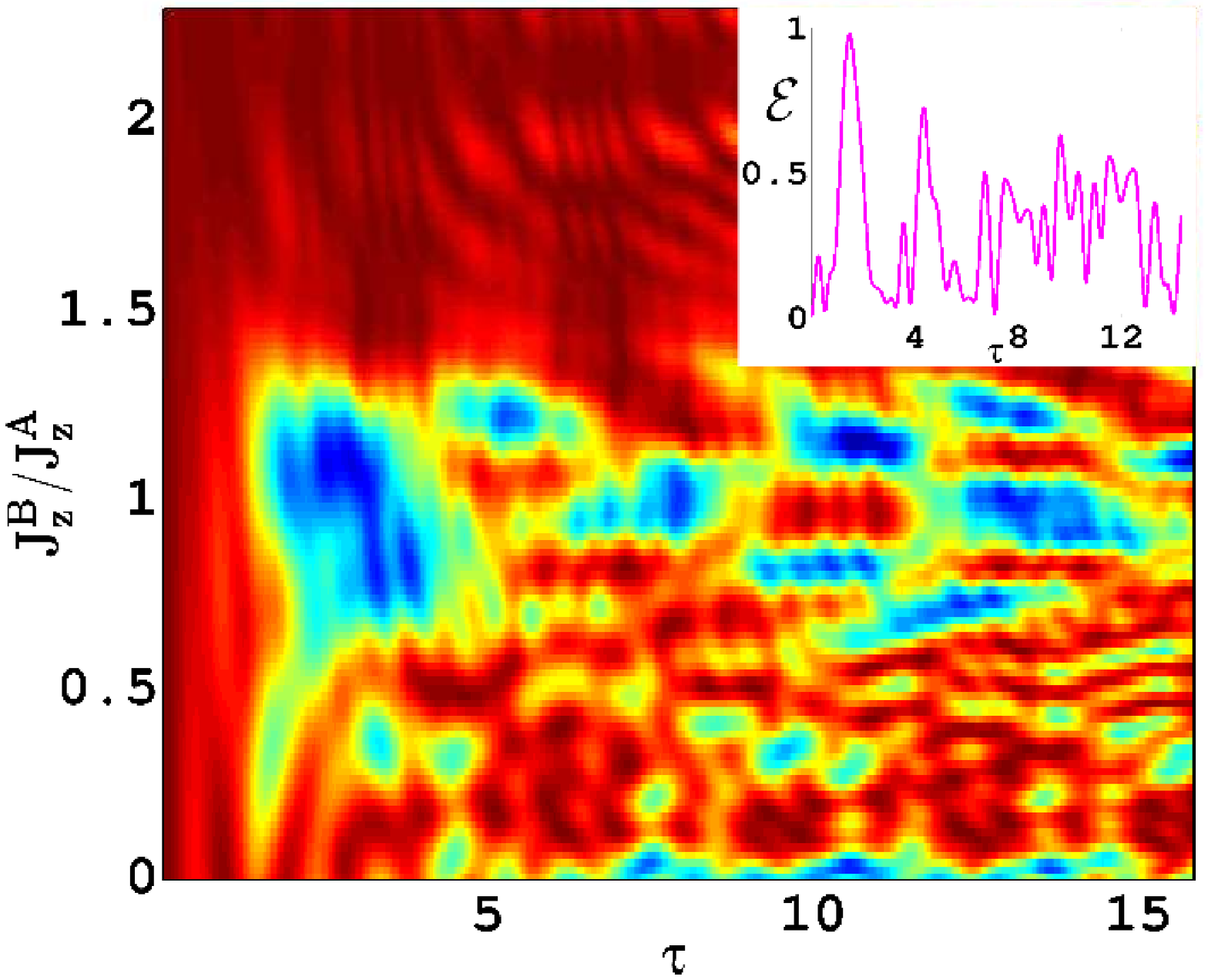}
\label{fig:RL_B005}
}
\caption{Rabi nutation of two coupled qubits in the geometry of Fig.~1(b) where both qubits are initialized in their $|0\rangle$ states. Qubit $A$ is driven at its level splitting $J_z^A$ with a Rabi frequency $\epsilon=0.3J_z^A$, and qubit $B$ is coupled to $A$ with a coupling constant $J_{AB}$. Time $\tau$ in periods of the free precession of $A$. Color scales are values of $S_z^A$ or $S_z^B$ between $-1$ and $1$. (a) $J_{AB}=0.02J_z^A$, color: nutation of ${\bf S}^B$. Insets: $S_x^B$ and $S_z^B$ for $J_z^B=0.3J_z^A$. (b) $J_{AB}=0.02J_z^A$. Color: nutation of ${\bf S}^A$. Inset: Entanglement $\mathcal E$ {\it vs} time $\tau$ for $J_z^B=0.3J_z^A$. (c) $J_{AB}=0.04J_z^A$, color: nutation of ${\bf S}^B$. Inset: Entanglement $\mathcal E$ {\it vs} time $\tau$ for $J_z^B=0.3J_z^A$. (d) The same, but for $J_{AB}=0.2J_z^A$. Inset: Entanglement $\mathcal E$ {\it vs} $\tau$ for $J_z^B=0.3J_z^A$. Notice developing of irregular behavior. See text for details.}
\label{fig:qubitB_RL_spin}
\end{figure*}

The RX qubit Hamiltonian for quantum dots in GaAs includes besides the dominant Heisenberg exchange two additional contributions from spin-orbit and hyperfine interactions \cite{Taylor, StepanSOIHF, Ladd, Klinovaja}. However, with $J_z^A,J_z^B,\epsilon\gg T_1^{-1},T_2^{-1}$ as in the experiments of Ref.~[\onlinecite{Medford2}], and a similar inequality for $J_{AB}$, we disregard dephasing and spin relaxation produced by these perturbations which is a subject of future work. Therefore, our results describe coherent dynamics of two entangled qubits. Under these conditions, a two-qubit system that starts at $t=0$ in a product state evolves as a pure two-qubit state $\vert\psi_{AB}\rangle$. Entanglement between qubits manifests itself through the Schmidt decomposition of $\vert\psi_{AB}\rangle$ in terms of single-qubit states \cite{NC}
\be
\vert\psi_{AB}\rangle=\sum_{i=1}^2c_i\vert\psi_i^A\rangle\vert\psi_i^B\rangle,
\label{Schmidt}
\ee
with $c_i\geq0$, $\sum_ic_i^2=1$. Entanglement can be quantified in terms of reduced density matrices $\rho_A={\rm Tr}_B\rho_{AB}$ and $\rho_B={\rm Tr}_A\rho_{AB}$, where $\rho_{AB}=\vert\psi_{AB}\rangle\langle\psi_{AB}\vert$, and partial traces are taken over the states of qubits $B$ and $A$, respectively. The von Neumann entropy associated with the reduced density matrices is a measure of entanglement in the system \cite{Bennett,Vedral}
\bea
{\mathcal E}&=&-{\rm Tr}(\rho_A\log_2\rho_A)=-{\rm Tr}(\rho_B\log_2\rho_B)\nonumber\\
&=&-\sum_ic_i^2\log_2c_i^2.
\label{entang}
\eea
The entanglement $\mathcal E$ takes its maximum value ${\mathcal E}=1$ for $c_1=c_2=1/\sqrt{2}.$

Below we present simulation results for the geometries of Figs.~1(b,c,d). We perform our simulation in the qubit space where the interaction Hamitonian is given in terms of Pauli matrices. The total Hamiltonian governing the qubit evolution is $\hat{H}_q + \hat{H}_{int}$. The ode15s function on MATLAB was used to evaluate the results. It is a variable-order solver that employs the Gear's method for solving differential equations and is suited for solving \textit{stiff} problems. Dynamics of spin ${\bf S}^B$ reflects the strength of coupling between qubits and intricacies of their joint dynamics, while $\mathcal E$ reflects the degree of quantum entanglement.

\subsection{Geometry of Fig.~1(b): R-L-coupling }
\label{sec:1b}

In this section we present the results of simulation for two qubits of Fig.~1(b) with capacitively coupled $R$ dot of qubit $A$ and $L$ dot of qubit $B$. Exact dynamics of both qubits and the effect of inter-qubit coupling are consistently taken into account. Qubit $A$ is driven by a resonant voltage $\epsilon\cos\omega t$ with the frequency $\omega=J_z^A$ applied between its $L$ and $R$ dots at $t=0$. At $t=0$, both dots are at the North poles of their Bloch spheres. Dimensionless time $\tau=J_z^At/2\pi$ is measured in periods of the free precession of qubit $A$. In Fig.~2(a), color reflects Rabi nutation of qubit $B$ driven by its coupling to qubit $A$ with $J_{AB}=0.02J_z^A$. The ratio $J_z^B/J_z^A$ is plotted along the vertical axis, and four resonances are seen. Three of these occur when $J_z^B$ is equal to $J_z^A$ and $J_z^A\pm\epsilon$, while fourth one when it equals to the Rabi frequency $\epsilon$. While only $J_z^A\pm\epsilon$ bands of the upper triplet are predicted by Eqs.~(\ref{eqSz})-(\ref{eqSy}), the strongest resonance corresponds to $J_z^B=J_z^A$; it corresponds to resonant transfer of excitation between A and B \cite{Mollow_triplet}. The modulation of $S_z^B$ may be understood as a two-step process, where, first, the oscillatory voltage at frequency $\omega$ mixes the states $S_z^A=1$ and $S_z^A=-1$ of qubit A, and then the term $J_{AB} \sigma_x^A \sigma_x^B$ leads to oscillations in $S_z^B$. Fig.~2(b) shows a similar plot but for the Rabi nutation of qubit $A$. It shows pronounced anomalies near all four frequencies discussed above. They provide direct demonstration of coupled dynamics of the qubits.

Insets to Fig.~2(a) demonstrate a $\pi/2$ nutation of qubit $B$ from the North pole to the equatorial plane and back to the North pole, and its fast precession about the $z$-axis. Observation of the nutation, by techniques of Ref.~[\onlinecite{Medford2}], should prove coupled dynamics of qubits and allow evaluating $J_{AB}$. 

Fig.~2(c) plotted for $J_{AB}=0.04J_z^A$ shows, in addition to four strong resonances, a number of weaker features which are due to anharmonicity of both qubits and are not seen for $J_{AB}=0.02J_z^A$.  They increase fast with $J_{AB}$, and for $J_{AB}=0.2J_z^A$, Fig.~2(d), patterns are highly irregular. The period in $\tau$ varies significantly with changes in $J_z^B/J_z^A$.

All panels of Fig.~2 were plotted under the condition of exact resonance $\omega=J_z^A$. To figure our whether this condition played an essential role in developing irregular pattern of Fig.~2(d), we recalculated the figure for $\omega=1.1J_z^A$ and found no drastic changes to it. While details changed, the basic pattern did not. This stability may be attributed to the width of the resonance of about $\epsilon$ that was rather large, equal to $\epsilon=0.3J_z^A$ in our simulations.
\begin{figure}[!hbtp]
\subfloat[]{
\includegraphics[width=0.21\textwidth]
{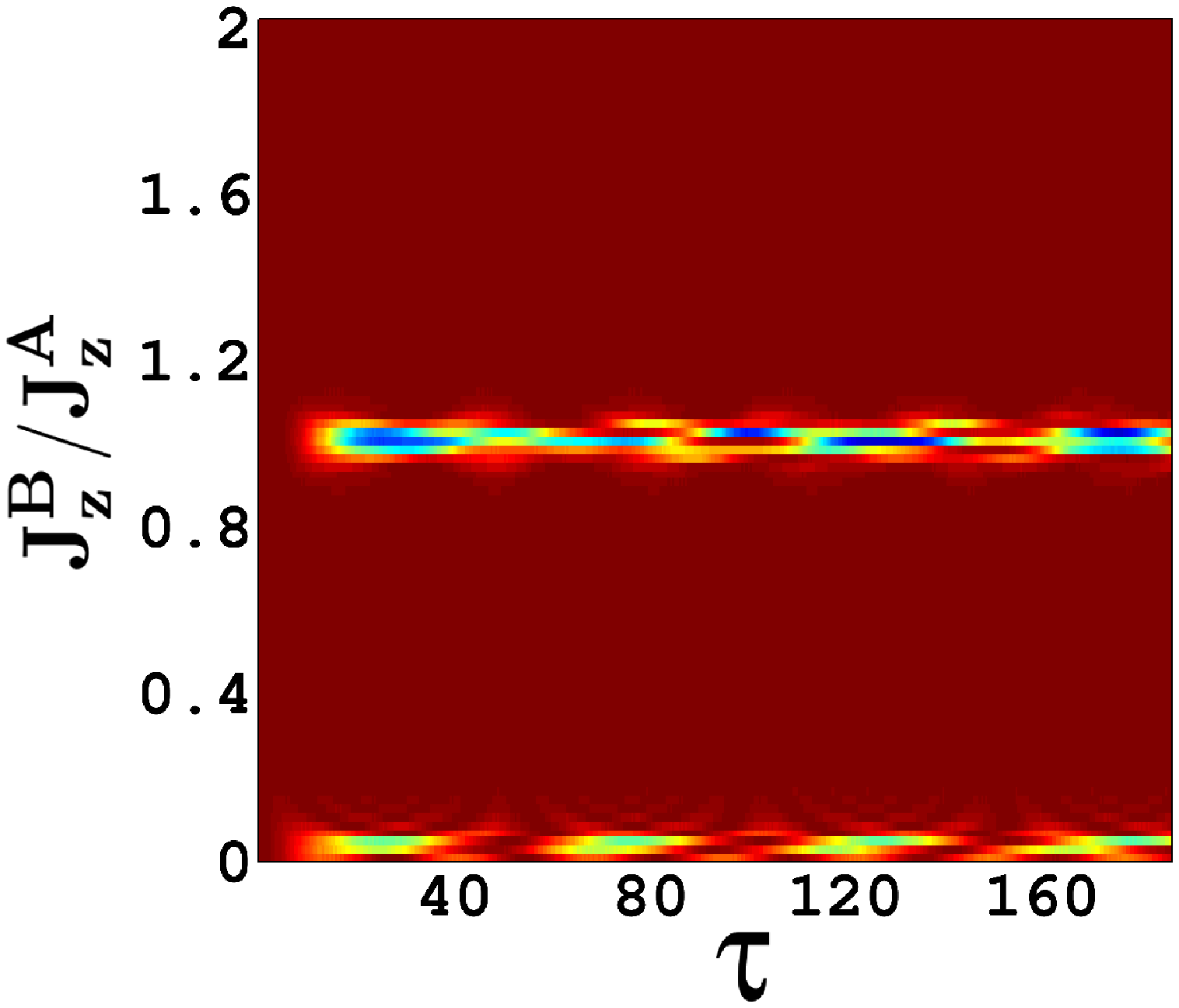}
\label{fig:RL_B_eA03Jab0005}
}
\subfloat[]{
\includegraphics[width=0.24\textwidth]
{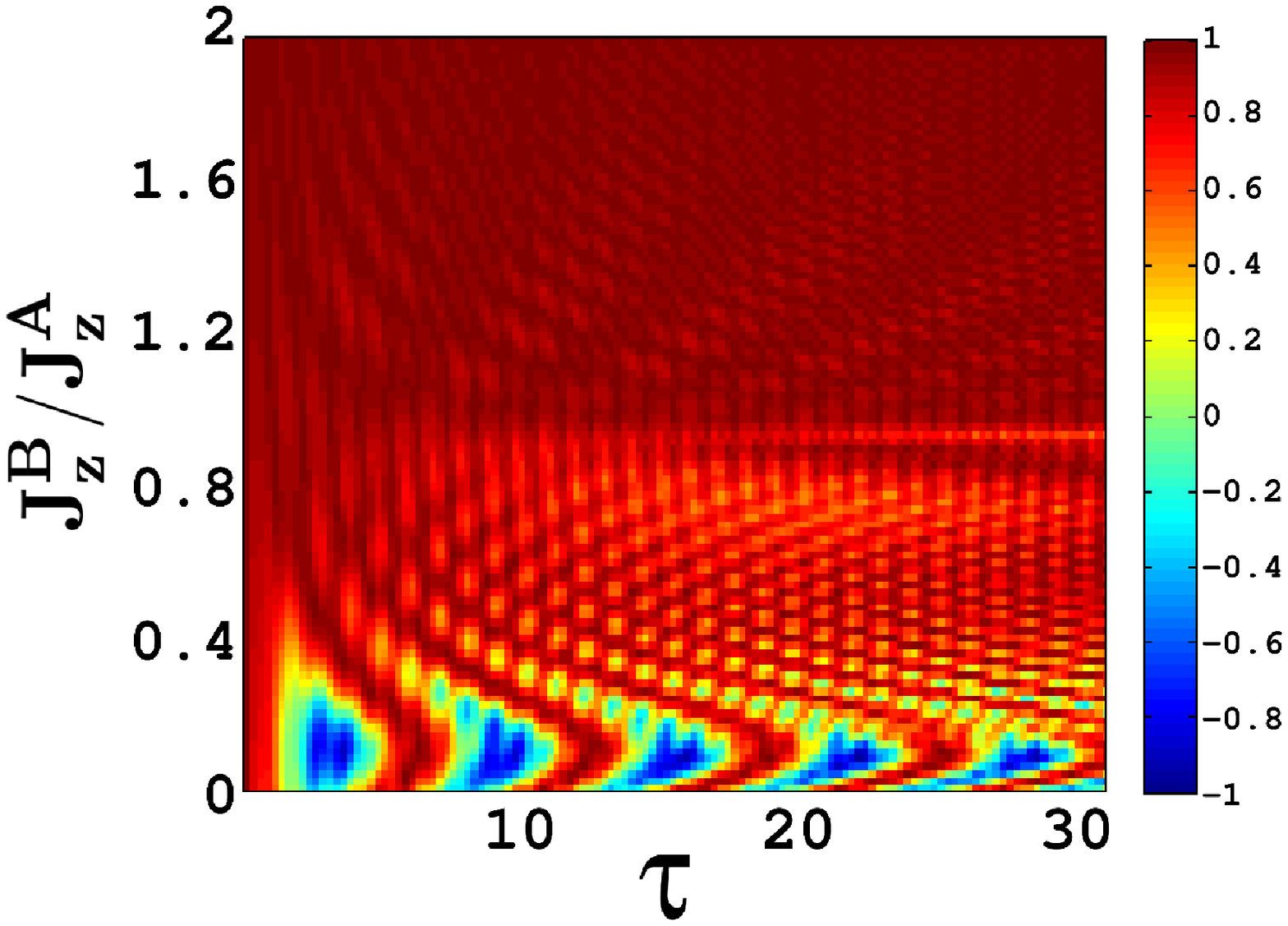}
\label{fig:RL_B_eA03Jab05}
}
\caption{Rabi nutation of two coupled qubits in the geometry of Fig.~1(b) where both qubits are initialized in their $|0\rangle$ states. Qubit $A$ is driven at its level splitting $J_z^A$ with a Rabi frequency $\epsilon=0.03J_z^A$, and qubit $B$ is coupled to $A$ with a coupling constant $J_{AB}$. Time $\tau$ in periods of the free precession of $A$. Color scale is the value of $S_z^B$ between $-1$ and $1$. (a) $J_{AB}=0.02J_z^A$. (b) $J_{AB}=0.2J_z^A$.}
\label{fig:qubitB_RL_eA}
\end{figure}

It is worth mentioning that a large Rabi frequency $\epsilon=0.3J_z^A$, while maintaining fast operation of qubit $A$, contributes to the multi-frequency irregular dynamics. With $\epsilon=0.03J_z^A$ and $J_{AB}=0.02J_z^A$, Rabi nutation of $B$ is dominated by a resonance at $J_z^B\approx J_z^A$ that developed by merging the $J_z^B=\omega$ and $J_z^B=\omega\pm\epsilon$ triplet of Fig.~2(a) into a single feature, see Fig.~3(a). It is accompanied by a weaker feature at $J_z^B\approx 0.04J_z^A$ which we attribute to slow nutation of qubit $B$ controlled by its $J_{AB}$ coupling to qubit $A$. Despite of weak driving, entanglement reaches the value
$\mathcal{E} \approx 0.95$ at $\tau \approx 50$ (for $J_z^A = J_z^B$). With $J_{AB}$ increasing to $J_{AB}=0.2J_z^A$, Figure 3(b)
, the $J_z^B \approx J_z^A$ resonance disappears while the low $J_z^B$ feature grows stronger, propagates to the region of higher $J_z^B\sim 0.2J_z^A$, and experiences pronounced oscillations of the Rabi nutation of $B$ with a period $\Delta\tau\approx5$. It may also be noted that the time scale of entanglement generation is unaffected by the decrease in Rabi frequency. Therefore, coupled nonlinear dynamics of two qubits shows high multiformity dependent on the specific choice of  parameter values.
\begin{figure*}[!hbtp]
\centering
\subfloat[]{
\includegraphics[width=0.35\textwidth]
{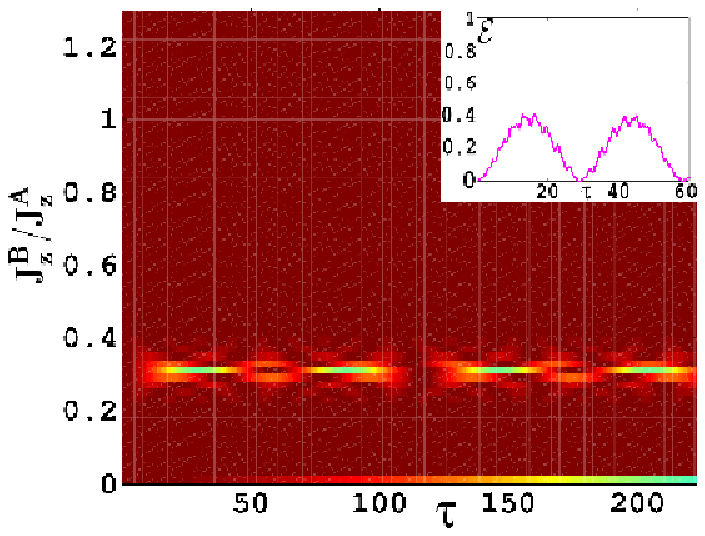}
\label{fig:ML_B0005}
}
\subfloat[]{
\includegraphics[width=0.36\textwidth]
{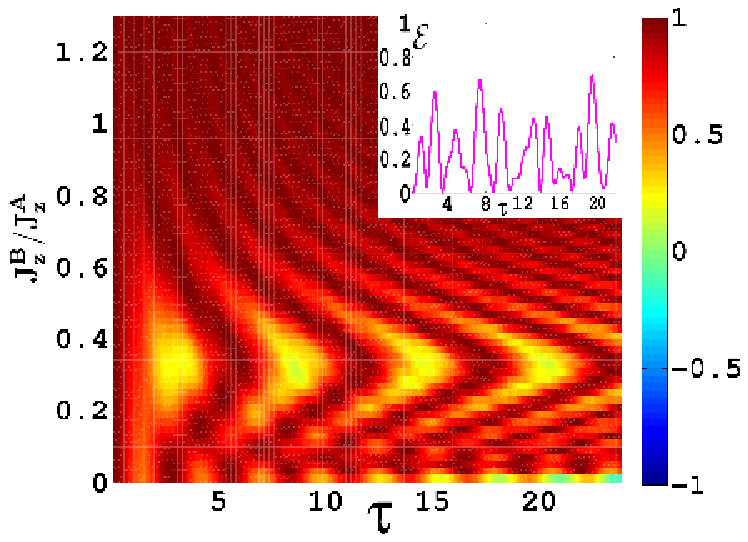}
\label{fig:ML_B005}
}
\\
\subfloat[]{
\includegraphics[width=0.42\textwidth]
{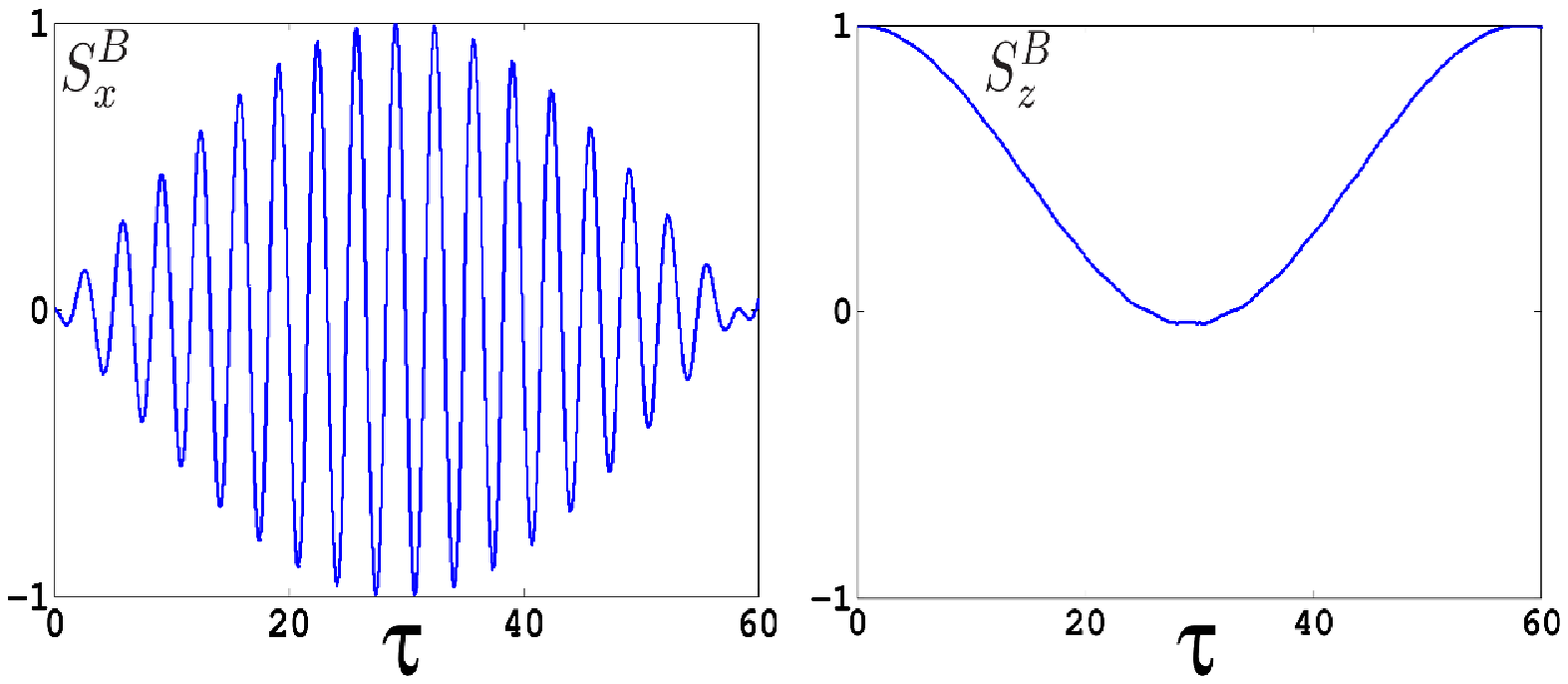}
\label{fig:ML_Btraject0005}
}
\subfloat[]{
\includegraphics[width=0.43\textwidth]
{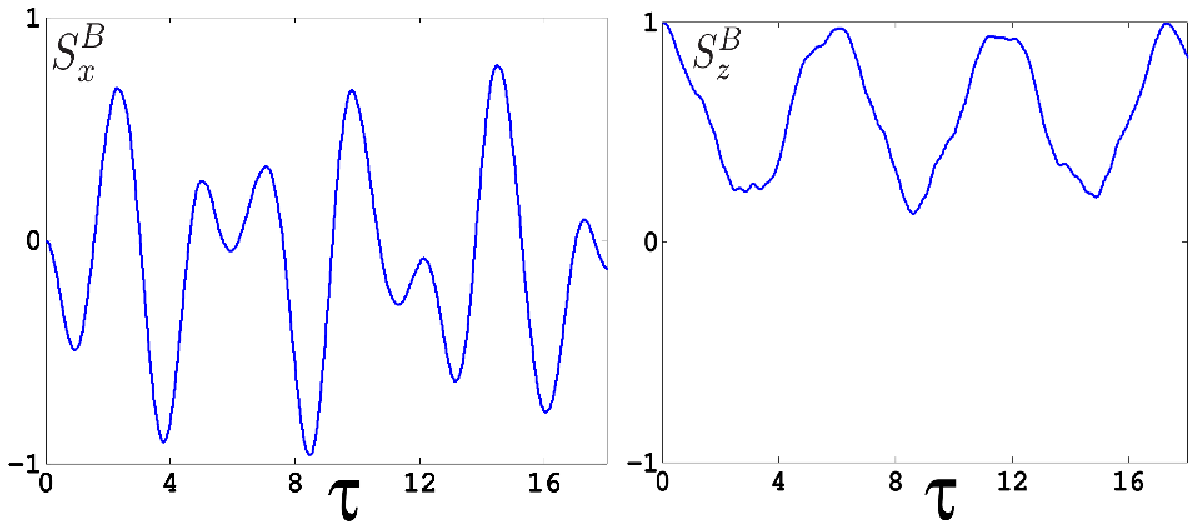}
\label{fig:ML_Btraject005}
}

\caption{ Rabi nutation of two coupled qubits in the geometry of Fig.~1(c). Qubit $A$ is driven at its level splitting $J_z^A$ with a Rabi frequency $\epsilon=0.3J_z^A$, and qubit $B$ is coupled to $A$ with a coupling constant $J_{AB}$. Time $\tau$ in periods of the free precession of $A$. Color scale in (a) and (b) is the value of Rabi nutation of ${\bf S}^B$ between $-1$ and $1$. (a) $J_{AB}=0.01J_z^A$. Inset: Entanglement $\mathcal E$ {\it vs} $\tau$ for $J_z^B=0.3J_z^A$. (b) $J_{AB}=0.1J_z^A$. Inset: Entanglement $\mathcal E$ {\it vs} $\tau$ for $J_z^B=0.3J_z^A$. (c) $S_x^B$ and $S_z^B$ for $J_z^B=0.3J_z^A$ and $J_{AB}=0.01J_z^A$. (d) $S_x^B$ and $S_z^B$ for $J_z^B=0.3J_z^A$ and $J_{AB}=0.1J_z^A$. See text for details.}
\label{fig:qubitB_ML_spin}
\end{figure*}

While the very fact of irregular behavior is not surprising having in mind the interplay of a number of frequencies and nonlinearity of the system, the fact that such a behavior sets-in in Fig.~2(d) at the values of the coupling constant as small as $J_{AB}\sim0.04J_z^A$ and at the time scale of a single nutation of qubit $B$ has important experimental implications. It suggests that establishing a well controlled entanglement of two qubits imposes restrictions on the coupling strength {\it not only from below but also from above}. With increasing $J_{AB}$, development of volatile behavior forestalls the increase in the operation speed. We note that early set in of volatile behavior is related to the presence of three frequencies in the responses $\delta n_\alpha$ of the $L$ and $R$ dots, and the geometries of Secs.~\ref{sec:1c} and \ref{sec:1d} show lesser volatility.

While the above discussion was focused on coupled dynamics of spins ${\bf S}^A$ and ${\bf S}^B$, insets to Figs.~2(b,c,d) display a purely quantum quantity, entanglement $\mathcal E$ between qubits. Remarkably, while nutation of $B$ in Fig.~2(c), for $\epsilon=0.3J_z^A$, 
reaches its maximum at $\tau_M\approx14$, the entanglement $\mathcal E$ is represented by a two-humped curve with the maxima on both sides of $\tau_M$ and a deep minimum near $\tau_M$. The same is true for insets to Fig.~2(a,b) with $\tau_M\approx32$. Such a behavior can be understood at a qualitative level because the nutation of B is entirely due to its coupling to A while nutation of A is primary due to the driving. Hence, it is nutation of B that reflects the entanglement. Near the first maximum of irregular nutation the dynamic of $S_z^B$ is slow, and this indicates that ${\bf S}^A$ and ${\bf S}^B$ are nearly decoupled, hence, entanglement $\mathcal E$ is small. On the contrary, fast change of $S^B_z$ indicates strong correlations between ${\bf S}^A$ and ${\bf S}^B$, hence, $\mathcal E$ is large. Inset to Fig.~2(d) demonstrates strong, while irregular, entanglement with a pronounced peak $\mathcal E\approx1$ at $\tau\approx1.5$, near the first maximum of the nutation of ${\bf S}^B$.

We note that despite quite a moderate coupling strength, $J_{AB}\sim0.04J_z^A$, strong nutation of $B$ and entanglement $\mathcal E$ develop at a scale of the time of $\tau\sim10$, still before the effects of dephasing and decoherence are expected to manifest themselves. Therefore, we envision existence of a considerable time scale where nonlinear dynamics dominates over dephasing.   

\subsection{Geometry of Fig.~1(c): M-L-coupling }
\label{sec:1c}

Coupled dynamics of qubits $A$ and $B$ in the geometry of Fig.~1(c) was calculated similarly to Sec.~\ref{sec:1b} but with a Hamiltonian of Eq.~(\ref{eqML}) rather than of Eq.~(\ref{eqRL}). Because coupling of qubit $B$ to qubit $A$ is represented by a single matrix $\sigma^A_z$, the input from $A$ is dominated by Rabi frequency $\epsilon$ as follows from  Eq.~(\ref{eqSz}). Correspondingly, response of qubit $B$ at $J_{AB}=0.01J_z^{A}$ is dominated by a single frequency $\epsilon=0.3J_z^A$ as seen from Fig.~4(a); satellites originating from nonlinearities are visible but very weak. Panel~4(c) demonstrates a $\pi/2$ nutation of qubit $B$ and its precession. Fig.~4(b) shows that patterns of the nutation of qubit $B$ remain basically regular even for $J_{AB}=0.1J_z^A$, with four oscillations occuring in the illustrated time period; however, detailed patterns show more features as seen from Fig.~4(d). This demonstrates that purification of the signal coming from the qubit $A$ preserves stability of entangled two-qubit dynamics and suppresses transition into irregular regime. Purification can be achieved by proper choice of geometry or by using cavities similar to Josephson qubits \cite{Josephson}.

Inset to Fig.~4(a,b) display time dependence of entanglement $\mathcal E$ for $J_z^B=0.3J_z^A$ where nutation of qubit $B$ experiences pronounced oscillations seen in the main panels. Similarly to inset to Fig.~2(c), two first maxima of $\mathcal E$ develop on both sides of the first peak of nutation. Therefore, for reasons explained in Sec.~\ref{sec:1b} peaks of ${\mathcal E}(\tau)$ correlate with extrema of $dS_z^B/d\tau$ rather than with extrema of $S_z^B$.

\subsection{Geometry of Fig.~1(d): M-M-coupling }
\label{sec:1d}

Fig.~5 presents dynamics of two coupled qubits in the geometry of Fig.~1(d). External resonant driving is applied to qubit $A$, and inter-qubit coupling is of $\sigma^A_z\sigma^B_z$ type and described by the Hamiltonian of Eq.~(\ref{eqMM}). This system possesses one integral of motion $\sigma_z^B$, and therefore the only dynamics of ${\bf S}^B$ is precession about the $z$ axis. Fig.~5 displays precession of qubit $B$ with the initial condition $S_x^B(t=0)=1$ and $S_z^A(t=0)=1$.  It is seen that with exclusion of small values of $J^B_z\alt\epsilon/J_z^A=0.3$, precession of $S^B_x$ is very regular. Gross features of the column structure of the plot are described by the phase modulation of Eq.~(\ref{eqSMM}) while gradual change of the color of columns should be attributed to backaction. Fast precession of $S_x^B$ is an intrinsic property of qubit $B$ controlled by $J_z^B$, but the columnar structure of Fig.~5 (distinctly reflected in $S_x^B(\tau)$ plot of the left inset) originates from the inter-qubit coupling and is controlled by  $J_{AB}$. 

Entanglement ${\mathcal E}(\tau)$ calculated for $J_z^B=0.3J_z^A$ is shown in the right inset of Fig.~5.  It has a shape of a dome on which oscillations correlated with the columnar structure in the main panel are superimposed. Remarkably, the maximum ${\mathcal E}\approx1$ is achieved simultaneously with the node in the oscillations of $S_x^B(\tau)$ seen in the left inset. Therefore, near the maximum of entanglement the precession of ${\bf S}^B$ becomes nearly frozen. 
\begin{figure}[!hbtp]
\includegraphics[width=3.0in, height=2.1in]{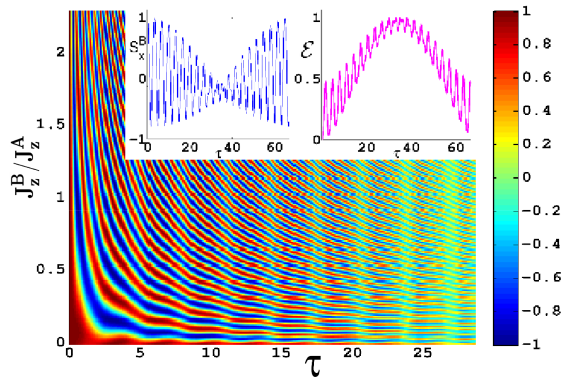}
\caption{Precession of qubit $B$ starting at $S_x^B=1$ in the geometry of Fig.~1(d). Qubit $A$ is driven at its level splitting $J_z^A$ with a Rabi frequency $\epsilon=0.3J_z^A$, and qubit $B$ is coupled to $A$ with a coupling constant $J_{AB}=0.05J_z^A$. Time $\tau$ in periods of the free precession of $A$. Color in the main panel and left inset: $S_x^B(\tau)$. Right inset: Entanglement ${\mathcal E}(\tau)$ for $J_z^B=0.3J_z^A$. See text for details.}
\label{fig:MM_B005}
\end{figure}

This observation deserves more detailed comments. As it has been mentioned below Eq.~(\ref{entang}), ${\cal E}=1$ is only possible for $c_1=c_2=1/\sqrt{2}$, hence, the reduced density matrices $\rho_A=\rho_B=\sigma_0/2$, where $\sigma_0$ is a unit matrix. Therefore, ${\bf S}^B={\rm Tr}(\rho_B {\bf \sigma}^B)=0$ because traces of Pauli matrices vanish, and similarly ${\bf S}^A=0$. For qubit B, $S_B^z$ is conserved and equals identically to $0$ due to initial conditions, $S_x^B\approx0$ at the maximum of $\cal E$ as seen in the left inset to Fig.~5, and it has been checked that $S_y^B\approx0$ there (not shown). We checked that the same is true for ${\bf S}^A$.

The results of this section, especially the data of Fig.~4(d), prove efficiency of the proposed protocol for producing entanglement of two qubits and evaluating coupling $J_{AB}$ between them. They also provide a warning against irregular dynamics in a two-qubit system, Fig.~2(d). It is typical of (i) the regime of fast operation, with comparable values of Rabi frequency and qubit level splitting, and (ii) interaction Hamiltonian with multiple non-commuting terms. 
In such a regime, efficient control of a two-qubit system bounds the magnitude of $J_{AB}$ not only from below but also from above. 

\section{Dependence on initial conditions} 
\label{sec:DIC}

In Sec.~\ref{sec:DE}, we investigated the effect of geometry on coupled two-qubit dynamics. The next problem is sensitivity of the dynamics to initial conditions. It is important both from the practical and conceptual point of view. First, it is difficult to control with high accuracy the initial state of a two-qubit system. Second, there is an open question whether dynamics of such a system is stable or chaotic at a large time scale. Indeed, dynamics of a two-qubit system is described by four nonlinear differential equations for two polar angles, $\theta_A$ and $\theta_B$, and two azimuths, $\varphi_A$ and $\varphi_B$, on their Bloch spheres. Because generically solutions of systems of nonlinear differential equations, beginning from three equations, are prone to deterministic chaos, such a behavior is also expected for two coupled qubits. To be specific, we have in mind a deterministic classical Lorenzian chaos \cite{Lorenz} rather than quantum chaos typical of mesoscopic systems \cite{Nakamura,Ullmo} or the quantum chaotic border on the number of qubits \cite{Georgeot}. Deterministic chaos manifests itself as a high sensitivity of solutions to small changes in the initial conditions.
 
To this end, we solved dynamic equations for a set of initial conditions as applied to the parameters of Fig.~2(d) where the dynamics was most irregular. The results, for $\epsilon=0.3J_z^A$, are presented in Fig.~6 for five values of $\theta_B^0=\theta_B(t=0)$, the polar angle of qubit $B$ at $t=0$. In all cases, $\varphi_B(t=0)=0$, and $\theta_A(t=0)=0$. It is seen that while each of the curves shows rich pattern, with a number of minima and maxima, these pattern are rather similar for all five curves plotted for values of $\theta_B^0$ ranging from $\theta_B^0=0$ to $\theta_B^0=48^0$. In particular, positions of all major features and even their magnitudes (in units of their initial values) remain very close. Therefore, we conclude that at least at this time scale (important for experiment) complexity of dynamics is dominated by nonlinearities of both qubits and their crosstalk and show no signatures of chaotic behavior.
\begin{figure}[!hbtp]
\includegraphics[width=3.0in]{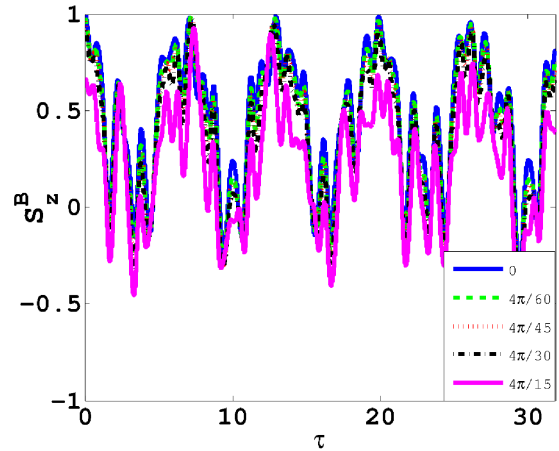}
\caption{Nutation of qubit $B$ for five values of $\theta_B^0$, the initial values of the polar angle $\theta_B$ of qubit $B$. All parameter values the same as in Fig.~2(d), $J_z^B=0.3J_z^A$. Complexities of dynamics are due to nonlinearity of both qubits and crosstalk between them. See text for details.}
\label{fig:RL_B005_initial_conditions}
\end{figure}

\section{Conclusions}
\label{sec:concl}

We proposed a protocol for producing entanglement between two qubits, and studied it analytically and numerically as applied to three geometries of capacitively coupled exchange only qubits. The protocol is based on driving qubit $A$ at its level splitting $J_z^A$ and reading the response of qubit $B$. We found that the protocol generates  entanglement when the coupling $J_{AB}$ between qubits is only about 1\% of $J_z^A$. We have also found that the patterns of the entanglement are highly sensitive to the double-qubit geometry and the mechanism of inter-qubit coupling. For fast qubits with Rabi frequency $\epsilon$ comparable to $J_z^A$, dynamics shows early switching to irregular behavior if qubit $B$ experiences a multi-frequency signal from the harmonically driven qubit $A$. Because anharmonicity is inherent in two-level systems, we compare various coupling schemes and select geometries allowing inhibiting early transition to irregular dynamics. To be specific, the geometry of Fig.~1(c) is more promising than the geometry of Fig.~1(b). We note that the requirement of stable entanglement of two qubits (i) may impose not only a lower but also an upper bound onto the inter-qubit coupling constant $J_{AB}$ and (ii) is more easily satisfied when inter-qubit coupling is dominated by a single frequency. Therefore, using cavities can facilitate entanglement not only through (i) enhancing the inter-qubit coupling and (ii) establishing connection between spatially separated qubits, but also through (iii) purification of the frequency spectrum of inter-qubit coupling. While our calculations were performed for capacitively coupled qubits, basic conclusions are applicable also to exchange coupled qubits because both charge and exchange densities are bilinear in the wave function amplitudes.

{\bf Acknowledgements.} Authors acknowledge useful discussions with J. Medford, C. M. Marcus, J. M. Taylor, and Ariel Amir. Research was supported by the Office of the Director of National Intelligence, Intelligence Advanced Research Projects Activity (IARPA), through the Army Research Office grant W911NF-12-1-0354  and by the NSF through the Materials Work Network program DMR-0908070. 

\appendix

\section{Derivation of the qubit Hamiltonian}
\label{appendix:Hq_deriv}

Basic properties of the exchange only three-electron triple-quantum-dot qubit are described by a Hubbard Hamiltonian
\begin{align}
\label{Hubbard}
\hat{H}=& \sum_{\langle ij \rangle, \sigma} t_{ij}\left( c^{\dagger}_{i, \sigma} c_{j, \sigma} + c^{\dagger}_{j, \sigma} c_{i, \sigma}  \right) \nonumber \\ &+ \sum_{i, \sigma} V_i c^{\dagger}_{i, \sigma} c_{i, \sigma} + \frac{U}{2}  
\sum_i n_i  \left(n_i-1\right),
\end{align}
where $t_{i, i+1}$ are hopping matrix elements between adjacent sites $i$ and $i+1$, $V_i$ is the electrostatic potential at the site $i$, and $U$ is the on-site Hubbard repulsion. The Hamiltonian $\hat H$  is acting in a Hilbert space spanned by four $S=1/2, S_z=1/2$ states. 
We choose two states of the ``charged" sector as 
\begin{align}
\label{eq:201states}
|201 \rangle &= c_{L\uparrow}^{\dagger}c_{L\downarrow}^{\dagger}c_{R\uparrow}^{\dagger}|\phi\rangle, \nonumber \\
|102 \rangle &= c_{L\uparrow}^{\dagger}c_{R\uparrow}^{\dagger}c_{R\downarrow}^{\dagger}|\phi\rangle, 
\end{align}
where $|\phi\rangle$ is the vacuum state. In the (111) region the states are 
\begin{align}
| \psi^{rl}_{111} \rangle &= \frac{1}{\sqrt{2}}\left( c_{L\downarrow}^{\dagger}c_{M\uparrow}^{\dagger}c_{R\uparrow}^{\dagger}-c_{L\uparrow}^{\dagger}c_{M\uparrow}^{\dagger}c_{R\downarrow}^{\dagger} \right)|\phi\rangle, \nonumber \\
| \psi^{lm}_{111} \rangle &= \frac{1}{\sqrt{2}}\left( c_{L\uparrow}^{\dagger}c_{M\downarrow}^{\dagger}c_{R\uparrow}^{\dagger}-c_{L\downarrow}^{\dagger}c_{M\uparrow}^{\dagger}c_{R\uparrow}^{\dagger} \right)|\phi\rangle, \nonumber \\
| \psi^{mr}_{111} \rangle &= \frac{1}{\sqrt{2}}\left( c_{L\uparrow}^{\dagger}c_{M\downarrow}^{\dagger}c_{R\uparrow}^{\dagger}-c_{L\uparrow}^{\dagger}c_{M\uparrow}^{\dagger}c_{R\downarrow}^{\dagger} \right)|\phi\rangle.
  \label{eq:111states}
\end{align}
The states in Eq. (\ref{eq:111states}) are not mutually orthogonal and only two of them are linearly independent. We choose the linear combinations that are symmetrical in the central basis 
\begin{align}
|111 \rangle &= \frac{1}{\sqrt{3}} \left( | \psi^{lm}_{111} \rangle + | \psi^{mr}_{111} \rangle \right) \nonumber\\
|\overline{111} \rangle &= | \psi^{rl}_{111} \rangle \label{eq:qubit}
\end{align}
We choose qubit states as $|0\rangle \approx |111\rangle$ and $|1\rangle \approx |\overline{111}\rangle$. In the basis $\left( |201\rangle, |111\rangle, |\overline{111} \rangle, |102\rangle \right)$ the total wavefunction is 
\begin{equation}
|\psi \rangle = v_L |201\rangle + v_0 |111\rangle + v_1 |\overline{111}\rangle + v_R |102\rangle
\label{eq:wavefn}
\end{equation}
In the above basis the eigenvalue equation takes the form 
\begin{equation}
\label{eq:Hmatrix}
\begin{pmatrix}
U+V & t_L\sqrt{\frac{3}{2}} & -\frac{t_L}{\sqrt{2}} & 0\\
t_L\sqrt{\frac{3}{2}} & 0 & 0 & -t_R\sqrt{\frac{3}{2}} \\
-\frac{t_L}{\sqrt{2}} & 0 & 0 & -\frac{t_R}{\sqrt{2}} \\
0 & -t_R\sqrt{\frac{3}{2}} & -\frac{t_R}{\sqrt{2}} & U-V  
\end{pmatrix} 
\begin{pmatrix}
v_L \\ v_0 \\ v_1 \\ v_R
\end{pmatrix}
=E
\begin{pmatrix}
v_L \\ v_0 \\ v_1 \\ v_R
\end{pmatrix},
\end{equation}
where components of the vector ${\bf{v}}=\left(v_L, v_0, v_1, v_R\right)^T$ are amplitudes of the states $|201 \rangle$, $|111 \rangle $, $|\overline{111} \rangle$, and $|102 \rangle$, respectively. 

This Hamiltonian can be projected into the $2\times2$ qubit space by eliminating the $(v_L,v_R)$ components
\begin{equation}
\label{eq:Hqubit} 
\begin{pmatrix}
-\frac{3}{2} \left( J_L + J_R \right) & \frac{\sqrt{3}}{2} \left( J_L - J_R \right)\\
\frac{\sqrt{3}}{2} \left( J_L - J_R\right) & -\frac{1}{2} \left( J_L + J_R \right)
\end{pmatrix}
\begin{pmatrix}
v_0 \\ v_1 
\end{pmatrix}
=E
\begin{pmatrix}
v_0 \\ v_1 
\end{pmatrix}
\end{equation}
where 
\be
J_L(E)=\frac{t_L^2}{U+V-E},\,\,J_R(E)=\frac{t_R^2}{U-V-E}.
\label{exint}
\ee
The qubit is operated under the conditions $t_L,t_R\ll U$, and therefore $E$-dependence of $J_L(E)$ and $J_R(E)$ is important mostly during the preparation and measurement cycles when $|V|\sim U$ \cite{Laird,Medford1}. In the central region of $|V|\ll U$, the energy is small, $E\sim t^2/U\ll U$, and can be omitted in the denominators of $J_L(E)$ and $J_R(E)$. Then, after shifting the origin of the energy in Eq.~(\ref{eq:Hqubit}) by $J_L+J_R$, we arrive at Eqs.~(\ref{eq1}) and (\ref{eq2}).

\section{Derivation of the interaction Hamiltonian}
\label{appendix:Hq_int}

For capacitively coupled qubits, the interaction Hamiltonian between excess charges on nearest-neighbor dots of the two qubits originate from the Coulomb interaction given by Eq.~\ref{eq:Hint}. The excess charge on each dot can be expanded in terms of the wavefunction amplitudes. For a symmetric system ($t_L=t_R=t$), the charges on the dots expressed in terms of the amplitudes $(v_0,v_1)$, after using the normalization condition $|v_0|^2 + |v_1|^2 = 1$, take the form
\begin{align}
\label{eq:QexcessLv}
\delta n_L &= \frac{t^2}{U^2} \left( 1+\frac{1}{2} |v_0|^2 - \frac{1}{2} |v_1|^2 -\frac{\sqrt{3}}{2} \left(v_0^* v_1 + v_1^* v_0 \right) \right), \\
\label{eq:QexcessRv}
\delta n_R &= \frac{t^2}{U^2} \left( 1+\frac{1}{2} |v_0|^2 - \frac{1}{2} |v_1|^2 +\frac{\sqrt{3}}{2} \left(v_0^* v_1 + v_1^* v_0 \right) \right), \\
\label{eq:QexcessMv}
\delta n_M &= -\frac{t^2}{U^2} \left( 2 + |v_0|^2 - |v_1|^2 \right). 
\end{align}
The charge \textit{operators} acting on the qubit space are given in terms of Pauli matrices. 
\begin{align}
\label{eq:QexcessL}
\delta \hat{n}_L &= \frac{t^2}{U^2} \left( \sigma_0 - \frac{\sqrt{3}}{2} \sigma_x +\frac{1}{2} \sigma_z\right), \\
\label{eq:QexcessR}
\delta \hat{n}_R &= \frac{t^2}{U^2} \left( \sigma_0 + \frac{\sqrt{3}}{2} \sigma_x +\frac{1}{2} \sigma_z\right), \\
\label{eq:QexcessM}
\delta \hat{n}_M &= -\frac{t^2}{U^2}  \left( 2\sigma_0 + \sigma_z \right).
\end{align}
In evaluating the dynamics of two coupled qubits the contribution of $\sigma_0$ to Eqs.~(\ref{eqRL})-(\ref{eqMM}) has been neglected. It only contributes to renormalize the coefficients of the single qubit terms in the Hamiltonian originating from the Coulomb interaction. We assume all such effects to be absorbed in the chosen values of $J_z$ and $J_x$. On subsituting the expressions from Eqs.~(\ref{eq:QexcessL})-(\ref{eq:QexcessM}) into Eq.~\ref{eq:Hint} gives us the form of the interaction Hamiltonian for the different geometries in Eqs.~(\ref{eqRL})-(\ref{eqMM}).


\begin{thebibliography}{99}
\bibitem{Aers_note}
Two-axis rotations were also achieved by employing hyperfine coupling near anticrossings of the lower qubit branch with $Q_{3/2}$ state \cite{Aers}.
\bibitem{Mollow_triplet}
The resonances $J_z^A$ and $J_z^A\pm\epsilon$ can be considered, in terms of laser physics, as an analogue of the Mollow triplet \cite{Mollow}.
\end{thebibliography}
\end{document}